\documentclass[12pt]{article}
\def\beq{\begin{equation}}
\def\eeq{\end{equation}}
\def\bea{\begin{eqnarray}}
\def\eea{\end{eqnarray}}
\def\nn{\nonumber}
\def\ba{\begin{array}}
\def\ea{\end{array}}

\begin{document}
\baselineskip16pt
\smallskip
\begin{center}
{\large \bf \sf
       Fermionic dual of one-dimensional bosonic particles \\ 
with derivative delta function potential}

\vspace{1.3cm}

{\sf B. Basu-Mallick$^1$\footnote{ 
e-mail address: bireswar.basumallick@saha.ac.in }
and Tanaya Bhattacharyya$^2$\footnote{e-mail address:
tanaya.bhattacharya@saha.ac.in } },

\bigskip

{\em $^1$Theory Group, \\
Saha Institute of Nuclear Physics, \\
1/AF Bidhan Nagar, Kolkata 700 064, India } \\

\bigskip

{\em $^2$Department of Physics, \\ 
Syamaprasad College, \\
5/B, R. Das Gupta Road, 
Kolkata 700 026, India } \\

\bigskip

\end{center}

\vspace {1.1 cm}
\baselineskip=16pt
\noindent {\bf Abstract }

We investigate the boson-fermion duality relation for the case of 
quantum integrable derivative $\delta$-function bose gas. In particular,
we find out a dual fermionic system with nonvanishing zero-range interaction 
for the simplest case of two bosonic particles with derivative 
$\delta$-function interaction. 
The coupling constant of this dual fermionic system becomes 
inversely proportional to the product of the 
coupling constant of its bosonic counterpart
and the centre-of-mass momentum of the corresponding eigenfunction.

\baselineskip=16pt
\vspace {.6 cm}

\vspace {.1 cm}

\newpage

\baselineskip=16pt
\noindent \section {Introduction }
\renewcommand{\theequation}{1.{\arabic{equation}}}
\setcounter{equation}{0}

\medskip

As is well known, 
 bosonic and fermionic theories in one spatial dimension
can often be related to each other through duality transformations. Such 
fermion-boson mappings for quantum integrable many-body systems 
play an important role in the study of their exact solutions.
A significant advancement in this direction has been done by
establishing the equivalence between one dimensional system of hard-core 
bosons and spinless free fermions [1].
Even though this equivalence between hard-core bosons and spinless free 
fermions was initially established only for the 
ground state of these systems, it can be shown that this
type of fermi-bose mapping holds true even for the excited 
states of impenetrable bosons as well as for all eigenstates of 
a quantum integrable bose gas interacting through two-body $\delta$-function
potential with arbitrary strength of the coupling constant [2,3].  
Indeed, it has been established that a 
novel fermionic model with non-vanishing zero range interaction can be mapped 
to the above mentioned bose gas interacting through $\delta$-function 
potential, where the coupling constants of these two models are
reciprocal to each other [3]. This fermion-boson equivalence immediately leads 
to the construction of  exact eigenfunctions of the dual fermionic 
model from the known eigenfunctions of the $\delta$-function bose gas. 

In this context it may be noted that,  
there exists another quantum integrable bosonic system 
interacting through derivative $\delta$-function potential 
with Hamiltonian given by 
\beq
H = -\sum_{j=1}^N\frac{\partial^2}{\partial x_j^2} + 2i
\xi ~\sum_{l<m}\delta(x_l - x_m)\Big( \frac{\partial}{\partial x_l}
+ \frac{\partial} {\partial x_m} \Big) \, ,
\label{a1}
\eeq
which can be obtained by projecting a certain type of derivative nonlinear
Schr\"{o}dinger (DNLS) field model to its the $N$-particle subspace [4-8]. 
Classical and quantum versions of such DNLS
model have found applications in different areas of physics like circularly
polarised nonlinear Alfven waves in plasma [9,10], 
quantum properties of optical
solitons in fibers [11] and in some chiral Luttinger liquids obtained from
Chern-Simons model defined in two dimensions [12,13]. Furthermore, 
it has been established that the ranges of coupling constant $\xi$,
 which allow the formation of bound states for the exactly solvable 
Hamiltonian (\ref{a1}), 
are determined through the Farey sequence in number theory [14]. 
Therefore, it should be interesting to explore whether their exists  an 
exactly solvable fermionic model which is dual to this 
bosonic system with derivative delta-function potential.
In Section 2 of this paper, we construct such a fermionic 
system with non-vanishing zero range interaction for the simplest
case of $N=2$.  For the sake of convenience, 
we  treat these systems in the infinite volume limit, i.e., when all 
particle coordinates can vary within the full range of the real axis.
It turns out that, for $N=2$ case,
the coupling constant appearing in the Hamiltonian of the dual fermionic system 
becomes inversely proportional  to the coupling constant of the bosonic Hamiltonian
and also to the centre-of-mass momentum of the corresponding eigenfunction.
In Section 3, we attempt to find out the dual fermionic model
 associated with the derivative $\delta$-function bose gas 
(\ref{a1}) for higher number of particles. 
Section 4 is the concluding section.

\vspace{1cm}

\noindent \section {Fermion-Boson duality for a system of two particles }
\renewcommand{\theequation}{2.{\arabic{equation}}}
\setcounter{equation}{0}

\medskip

Before dealing with the case of derivative $\delta$-function potential, let 
us briefly discuss how the bosonic two particle system with 
$\delta$-function potential can be mapped to a fermionic system with 
short-range interaction [3]. At first, the Hamiltonian of the bosonic
two particle system with $\delta$-function potential is reduced 
to a single particle Hamiltonian depending only on the relative coordinate.  
Although such a Hamiltonian preserves the continuity of the wavefunction,
it induces a discontinuity in the spatial derivative of the 
wavefunction. It turns out that the fermionic counterpart of this 
bosonic wavefunction behaves in exactly opposite way; 
it has a discontinuity in the
wavefunction itself but not in its spatial derivative. By applying the method 
of self-adjoint extension, it is possible 
to construct a short-range potential which induces this type of discontinuity
in the wavefunction and which, therefore, yields the Hamiltonian of 
the dual fermionic system. In the following, we shall use this approach 
to find out the dual fermionic system corresponding to
the bosonic two particle system with derivative 
$\delta$-function potential.

For $N=2$, the Hamiltonian (\ref{a1}) containing derivative $\delta$-function 
interaction may be written as   
\beq
H = - \frac{\partial^2}{\partial {x_1}^2}- \frac{\partial^2}{\partial {x_2}^2}
+ 2i\xi \, \delta (x_1-x_2)
\Big(\frac{\partial}{\partial x_1} + \frac{\partial}{\partial
x_2} \Big) \, .
\label {b1}
\eeq
Let us denote the bosonic wavefunction corresponding to
 this two-particle  system 
by $\psi_+(x_1, x_2)$, which satisfies the condition $\psi_+(x_1, x_2) = 
\psi_+(x_2, x_1)$.
With the help of the relative and centre of mass coordinates, i.e. 
$x=x_2-x_1$ and $X=\frac{1}{2}(x_1+x_2)$ respectively,
the Schr${\rm {\ddot o}}$dinger equation 
$H \psi_+ = E \psi_+$ can be expressed as 
\beq
-2\frac{\partial^2\psi_+}{\partial x^2} - \frac{1}{2}\frac{\partial^2\psi_+}
{\partial X^2} + 2i\xi\delta(x_1 - x_2)\frac{\partial \psi_+}{\partial X} 
= E \psi_+ \, .
\label{b2}
\eeq
It may be observed that the Hamiltonian (\ref{b1}) commutes with the 
total momentum operator $P$ given by 
\beq
P= -i \frac{\partial}{\partial x_1} - i\frac{\partial}{\partial x_2}\, .
\label{b3}
\eeq
Consequently, $\psi_+$ might be chosen as a simultaneous eigenstate of the
Hamiltonian (\ref{b1}) and the momentum operator (\ref{b3}). So  
we write $\psi_+$ in the form 
\bea
\psi_+=e^{iKX}\phi_+(x) \, ,
\label{b4}
\eea
where $\phi_+(x) =
\phi_+(-x)$ and
$P\psi_+ = K\psi_+ \, $. 
Substituting the expression of $\psi_+$ (\ref{b4}) in Eq.(\ref {b2}), 
we obtain
\beq
-\frac{\partial^2 \phi_+(x)}{\partial x^2} - K\xi\delta(x)\phi_+(x)= E_r \, 
\phi_+(x)
\label{b5}
\eeq
where $E_r= \frac{E}{2} - \frac{K^2}{4}$. 
Note that the  above equation can be 
written in the form 
$H_r \phi_+(x)= E_r \,\phi_+(x),$
where $H_r$ is given by 
\beq
H_r= - \frac{\partial^2}{\partial x^2} - K\xi\delta(x) \, .
\label{b5a}
\eeq
It is interesting to observe that, in contrast to the case of two particles 
interacting through $\delta$-function potential [3],  coupling constant
of the Hamiltonian (\ref{b5a}) now depends on the
centre-of-mass momentum of the two-particle wavefunction. In the following, 
it will be shown that the fermionic dual of $H_r$ also explicitly depends 
on this centre-of-mass momentum.

By integrating Eq. (\ref {b5}) within a small interval $[-\epsilon, \epsilon \, ]$
 and taking $\epsilon \rightarrow 0$ limit, we obtain 
\beq
\frac{\partial \phi_+}{\partial x}\Big|_{x\rightarrow +0} -~ 
\frac{\partial \phi_+}{\partial x}\Big|_{x\rightarrow -0} = - K \xi \phi_+
\Big|_{x=0}\, .
\label{b6}
\eeq
Applying the symmetry relation $\phi_+(x)=\phi_+(-x)$, it is easy to see 
that while the spatial derivative of $\phi_+(x)$ changes its sign around $x=0$,
the wavefunction itself is continuous around this point:
\beq
\phi_+(0_+) = \phi_+(0_-)\,;~~~~~\phi_{+}^{\prime}(0_+) =
-\phi_{+}^{\prime}(0_-)\,.
\label{b7}
\eeq
where
$\phi_+(\,0_{\pm}\,)= \lim_{a\rightarrow 0}\,\phi_+(\,\pm \,a\,)$ and 
$\phi_+^{\prime}(\, 0_{\pm}\,)= \lim_{a\rightarrow 0}\,\phi_+^{\prime}
(\,\pm \,a\,) $. 
Now, by simultaneously using Eqs. (\ref{b6}) and (\ref{b7}), 
we get the boundary 
conditions on the bosonic wavefunction as 
\beq
 \phi_+^{\prime}(0_+) = -  \phi_+^{\prime}(0_-)
= -\frac{K\xi}{2} \, \phi_+(0_+)= -\frac{K\xi}{2} \, \phi_+(0_-) \, .
\label{b8}
\eeq
Next, we define another wavefunction depending on $x$ as 
\beq
\phi_-(x)= [\,\theta(x)-\theta(-x)\,]\,\phi_+(x)\, ,
\label{b9}
\eeq
where $\theta(x) = 1$ when $x >0$ and $\theta (x) = 0$ when $x < 0$.
It is easy to see that $\phi_-(x)$ satisfies the 
symmetry relation: $\phi_-(-x)= -\,\phi_-(x)$. Hence, if we consider a
state like  $\psi_-(x)=e^{iKX}\phi_-(x)$, 
it will represent a  two particle fermionic
wavefunction. Using Eqs. (\ref{b8}) and (\ref{b9}), we find that the 
wavefunction $\phi_-(x)$ and its spatial derivative 
satisfy the following conditions around $x=0$:
\beq
-\phi_-(\,0_+)= \phi_-(\,0_-) = \frac{2}{K\xi}\, 
\phi_-^{\prime}(\,0_+) = \frac{2}{K\xi}\, \phi_-^{\prime}(\,0_-)\, .
\label{b10}
\eeq
Hence, the fermionic wavefunction $\phi_-(x)$ is discontinuous 
around $x=0\,$,
while its spatial derivative is continuous around this point. 

Even though in elementary quantum mechanics one does not usually encounter 
potentials leading to the discontinuity of the wavefunctions,
by using the method of self-adjoint extension it is possible to 
show that short-range potentials can generate discontinuity 
for both the wavefunction and
its spatial derivative [15-19]. Denoting the wavefunction associated with 
such short-range potential by $\phi(x)$ and using notations 
like before, we can express the discontinuity around the point $x=0$ as 
\beq
\phi^\prime (0_+) + \alpha \phi^\prime (0_-) = -\beta \phi(0_-) , ~~
\phi(0_+) + \gamma \phi(0_-) = -\delta \phi^\prime(0_-) \, ,
\label{b11}
\eeq
where $\alpha$, $\beta$, $\gamma$ and $\delta$ are arbitrary real numbers
satisfying the constraint 
\beq
\alpha \gamma - \beta \delta = 1 \, . \nn
\label{b12}
\eeq
The short range potential which gives rise to the discontinuity
(\ref{b11}) may be expressed in the form [19]
\beq
\chi(x;\, \alpha,\, \beta, \,\gamma, \,\delta) = \lim_{a \rightarrow
+0}\, [\,u_-\delta(x+a) + u_0 \delta(x) + u_+ \delta(x-a)\, ] \nn
\label{b13}
\eeq
where
\beq 
u_+(a) = -\frac{1}{a} + \frac{\alpha - 1}{\delta}, ~~ 
u_-(a) = -\frac{1}{a} + \frac{\gamma - 1}{\delta}, ~~ 
u_0(a) = \frac{1- \alpha\gamma}{\beta a^2} \, .
\label{b14}
\eeq
Note that if we choose the values of the free parameters as 
$\alpha = -1,$ $ \beta =
0,$  $\gamma = -1,$  and $\delta = \frac{4}{K\xi}$,
and also assume that $\phi(x)$ satisfies the fermionic 
exchange relation $\phi(-x)=-\phi(x)$, then the general form of discontinuity 
relation (\ref{b11}) reduces to Eq.(\ref{b10}). Consequently, by substituting 
the above mentioned values of the parameters in 
Eq.(\ref{b13}),  we obtain the explicit form of 
short range potential which introduces the discontinuity (\ref{b10})
in the fermionic wavefunction: 
\bea
\epsilon\,(\,x\,;\,\frac{1}{K\xi}\,) &\equiv
&\chi(x;\, -1,\, 0,\, -1,\, \frac{4}{K\xi}) \nn \\
&= &\mathop {\lim}\limits_{a\to
+0}\Big(-\frac{K\xi}{2}-\frac{1}{a}\,\Big)\,\Big\{\, \delta(x+a) + 
\delta(x-a)\,\Big\} \, .   
\label{b15}
\eea
Thus the fermionic dual to the bosonic Schr${\rm {\ddot o}}$dinger equation
(\ref{b5}) is given by  
\beq
-\frac{\partial^2\phi_-(x)}{\partial x^2} + \epsilon\,\Big(\,x\,;\,
 \frac{1}{K\xi}\,\Big)\,\phi_-(x)= E_r \,\phi_-(x) \, .
\label{b16}
\eeq
Now if we construct a two particle fermionic wavefunction like   
$\psi_-(x) = e^{iKX}\phi_-(x)$,  it will evidently satisfy
Schr${\rm {\ddot o}}$dinger equation of the form
\beq
-\frac{\partial^2\psi_-}{\partial x_1^2} -\frac{\partial^2\psi_-}{\partial
x_2^2}+ 2\epsilon\,(x_2-x_1; \frac{1}{K\xi})\psi_- = E \psi_-  \, .
\label{b17}
\eeq
Hence the two-particle fermionic system, which is dual to the two-particle 
bosonic system (\ref{b1}) with derivative $\delta$-function interaction,
is described by the Hamiltonian
\beq
H=-\frac{\partial^2}{\partial {x_1}^2} - \frac{\partial^2}{\partial{x_2}^2}
+2\epsilon\,\Big(\,x_2-x_1\,;\, \frac{1}{K\xi}\,\Big)\, .
\label{b18}
\eeq

It is interesting to observe that, in contrast to the 
case of $\delta$-function interaction, 
the fermionic Hamiltonian (\ref{b18}) explicitly depends on the 
centre-of-mass momentum. More precisely, the coupling 
constant of this fermionic Hamiltonian becomes inversely proportional 
to the product of the coupling constant of the corresponding bosonic system and 
its centre-of-mass momentum.  Thus, for the present case, 
the boson-fermion duality 
can be established only if we fix the centre-of-mass momentum 
of the two-particle bosonic system.  
In this context it may be noted that,  the binding energy
of the bosonic bound states does not have any momentum dependence 
for the case of $\delta$-function interaction [20]. 
This result is consistent with the fact that
the Hamiltonian of the corresponding dual fermionic system does not depend 
of the variable $K$ [3]. On the other hand, for the case of  
derivative $\delta$-function interaction, the  binding energy of the bosonic 
bound state depends on the total momentum of the system in a nontrivial way 
[14]. Therefore, the appearance of variable $K$
in the dual Hamiltonian (\ref{b18}) is rather natural, which merely
ensures  that the binding energy of two-particle fermionic bound state 
would also depend on the centre-of-mass momentum.

\noindent \section 
{Boson-fermion duality for higher number of particles
}

\renewcommand{\theequation}{3.{\arabic{equation}}}
\setcounter{equation}{0}

\medskip
Here our aim is to generalize the treatment of the earlier section to  
find out the dual fermionic model of the
derivative $\delta$-function bose gas (\ref{a1}) for $N\geq 3$. 
For the case of $\delta$-function bose gas, such a generalization 
is quite straightforward because this many-body problem 
can be effectively reduced to a set of two-body problems,
 whose relative coordinates 
play the key role. Consequently,  the boundary value 
problem associated with the $\delta$-function potential can be written 
through a set of equations, each of which depends on the 
partial derivative of the wavefunction with respect to only  
one relative coordinate [21,3]. 
Therefore, by applying again the method of self adjoint 
extension to the case of short range potential depending on a single 
relative coordinate, it is possible find out the dual fermionic
model for this many-body system. However, it has been found recently  
that derivative $\delta$-function bose gas exhibits a few 
nontrivial features for $N\geq 3$ [14]. For example, while
the chirality property of the corresponding classical solitons 
is preserved in the quantum theory for $N=2$, this chirality property is 
broken in the quantum theory for $N\geq 3$. Therefore, it is 
important to check whether the boundary value
problem associated with derivative $\delta$-function bose gas can in general 
be written through a set of equations, each of which would depend on the 
partial derivative of only one relative coordinate. 
For the sake of convenience,   
in the following we shall investigate this problem for
derivative $\delta$-function bose gas containing three particles. 
  
For the case $N=3$, the Hamiltonian (\ref{a1}) may be written as 
\bea
H &=& - {\bigtriangledown}^2
\,+\, V(x_{12})\,\Big(\,\frac{\partial}{\partial x_1}
+ \frac{\partial}{\partial x_2}\,\Big)  \nn \\ 
&&+\,V(x_{23})\,\Big(\,\frac{\partial}
{\partial x_2} +
\frac{\partial}{\partial x_3}\,\Big) +\,V(x_{13})\,\Big(\,\frac{\partial}
{\partial x_1} +
\frac{\partial}{\partial x_3}\,\Big)\, ,
\label{c1}
\eea
\medskip
where 
${\bigtriangledown}^2 \equiv \frac{\partial^2}{\partial x_1^2} + 
\frac{\partial^2}{\partial x_2^2} +
\frac{\partial^2}{\partial x_3^2}~$ and
$~V(x_{ij}) \equiv 2i\xi \delta(x_i - x_j)$.
It is evident that this Hamiltonian commutes with the total momentum operator given by
\beq
P= -i \,\Big(\frac{\partial}{\partial x_1} + \frac{\partial}{\partial x_2} +
\frac{\partial}{\partial x_3}\Big)\, .
\label{c2}
\eeq 
Therefore, the bososnic wavefunction $\psi_+(x_1,x_2,x_3)$ of  
three-body derivative $\delta$-function bose gas can be chosen as 
a simultaneous eigenstate of the Hamiltonian (\ref{c1}) and the momentum
operator (\ref{c2}). Hence we write $\psi_+(x_1,x_2,x_3)$ 
in a product form like 
\bea
\psi_+ (x_1,x_2,x_3)= e^{iK(x_1 + x_2 +x_3)}\phi_+  \, , 
\label{c2a}
\eea
where $\phi_+$ depends only on the relative coordinates associated 
with three particles. Since $\phi_+$ 
remains invariant under the simultaneous
translation of all coordinates, we obtain $P\phi_+ = 0$. Thus, due to
Eq.(\ref{c2a}) it follows that,   
$\psi_+$ would be an eigenstate of the total momentum operator 
with eigenvalue $3K$. Substituting  the form (\ref{c2a}) of $\psi_+$  
to the Schr\"{o}dinger equation given by $ H\psi_+ = E\psi_+$, we obtain 
\bea
&&-{\bigtriangledown}^2\phi_+ + 2iK\big\{\,V(x_{12}) + V(x_{23}) 
+
V(x_{13})\,\big\}\phi_+ \nn \\
&&~~~~~~~~~~~~~- \Big\{\,V(x_{12})\,\frac{\partial}{\partial x_3} +
V(x_{23})\,\frac{\partial}{\partial x_1} + V(x_{13})\,\frac{\partial}
{\partial
x_2}\,\Big \}\phi_+ = (E-3K^2)\phi_+\,. \nn \\
\label{c3}
\eea 

Let us now consider the situation where the first and second particles 
 move towards each other such that their centre of mass
remains in a fixed position, and the third particle remains stationary far away from  
these two particles.
Hence, we make a transformation of the coordinates like 
$$\big\{x_1, x_2, x_3 \big\} \Longrightarrow \big\{x= x_1-x_2,~
X= x_1+x_2,~x_3 \big\}\, ,$$
and try to find the boundary condition on  
$\phi_+$ following from Eq.(\ref{c3}) when only the coordinate $x$ varies 
around $x=0$ and the other two coordinates $X$ and $x_3$ have some fixed 
values.  Under the above mentioned change of coordinates, 
the ${\bigtriangledown}^2\phi_+$ term in
Eq.(\ref{c3}) can be expressed as a sum of three terms $\frac{\partial^2\phi_+}{\partial
x^2}$, $\frac{\partial^2\phi_+}{\partial X^2}$ and 
$\frac{\partial^2\phi_+}{\partial x_3^2}$ with different coefficients.
Since only the partial derivative of $\phi_+$ with respect to $x$ has a 
discontinuity around $x=0$,  the term $\frac{\partial^2\phi_+}{\partial x^2}$
will only give a non-zero contribution if we 
integrate  ${\bigtriangledown}^2\phi_+$   
within a small interval $[-\epsilon, +\epsilon\, ]$ and finally take 
$\epsilon \rightarrow 0$ limit.  Furthermore, when the potential
energy terms in Eq.(\ref{c3}) are 
integrated with respect to $x$ within the small interval specified above,
only those terms would contribute which have a 
$\delta$-function type or derivative $\delta$-function type singularity at $x=0$.
Consequently, by integrating (\ref{c3}), we obtain
\bea
\lim_{\epsilon \rightarrow 0} \left \{
-2\int_{-\epsilon}^{\epsilon}\frac{\partial^2\phi_+}{\partial x^2}\,dx +
2iK\int_{-\epsilon}^{+\epsilon}V(x_{12})\, \phi_+\, dx -
\int_{-\epsilon}^{\epsilon} V(x_{12})\frac{\partial \phi_+}{\partial x_3}\,dx 
\right \} = 0\, , 
\nn 
\eea
which reduces finally to 
\bea
\frac{\partial \phi_+}{\partial x}\Big|_{x=+\epsilon} - \frac{\partial
\phi_+}{\partial x}\Big|_{x= - \epsilon} = -2\xi K \phi_+\Big|_{x=0} -
i\xi \frac{\partial \phi_+}{\partial x_3}\Big|_{x=0}\, .
\label{c3a}
\eea 
It is worth noting that, in contrast to $N=2$ case, 
 this boundary condition is not expressed
only through the partial derivative with respect 
to the relative coordinate $x$. In fact, due to the presence of the last term
in the r.h.s. of Eq.(\ref{c3a}), the form of this equation differs from that of  
Eq.(\ref{b6}) which has been derived in the earlier section.

However, we can still proceed with the boundary condition (\ref{c3a})
 in the same way as has been done with the boundary condition (\ref{b6}) for $N=2$ case, 
provided it is possible to choose $\phi_+$ in some particular form 
 satisfying the condition 
\bea
\frac{\partial\phi_+}{\partial x_3} \Big|_{x=0} = \lambda \phi_+ \Big|_{x=0}\,  ,
\label{c4}
\eea
where $\lambda $ is an arbitrary constant. In this context it should be noted that,
the exact $N$-particle eigenfunction for derivative $\delta$-function
bose gas can be obtained through the Bethe ansatz [4,5,14]. In the region 
$x_1<x_2<\cdots <x_N$, such $N$-particle eigenfunctions may be written as 
\beq
\psi_N(x_1, x_2 , \cdots , x_N)= \sum_\omega\left (\prod
_{l<m}\frac{A(k_{\omega(m)},k_{\omega(l)})}{A(k_m,k_l)}\right)
\exp \, \{ i (k_{\omega(1)}x_1 + \cdots + k_{\omega(N)} x_N ) \} 
\, ,
\label{c5}
\eeq
where  $k_m$s are all distinct wave numbers, 
$\big(\omega(1), \omega(2), \cdots , \omega(N) \big)  $ 
represents a permutation of the numbers (1,2,....N)  
and $\sum_{\omega}$ implies summing over all such 
permutations. In the expression (\ref{c5}), 
the `matching coefficient' $A(k_m,k_l)$ is given by
\bea
A(k_m,k_l)=
\frac{(k_m-k_l)+i\xi(k_m+ k_l)}{(k_m-k_l)} \,  .\nn 
\eea
So, our next aim is to verify whether the condition (\ref{c4}) is satisfied 
for the 3-particle Bethe wavefunction associated with the 
derivative $\delta$-function bose gas. It is easy to check that, 
for the case $N=3$, the Bethe wavefunction (\ref{c5}) 
within the region $x_1<x_2<x_3$ can be written in  
the form of Eq.(\ref{c2a}),  where
$ K=\frac{k_1 + k_2 + k_3}{3}$,
\beq 
\phi_+ = \sum_{i=1}^{6} \phi_i,
\label{c5a}
\eeq
and $\phi_i$s' are explicitly given by
\bea
&&\phi_1 \equiv \phi_1(k_1, k_2, k_3)= e^{\frac{i}{3} \{x_1(2k_1-k_2-k_3) 
+ x_2 (2k_2-k_1-k_3)
+ x_3 (2k_3- k_1 - k_2)\}}\,,  \nn \\ 
&&\phi_2 = \frac{A(k_1, k_2)}{A(k_2, k_1)}\phi_1(k_1, k_2, k_3)\,,  \nn \\
&&\phi_3 = \frac{A(k_1,k_2) A(k_1, k_3)}{A(k_2, k_1)A(k_3, k_1)}
\phi_1(k_2, k_3, k_1)\,,  \nn \\
&&\phi_4 = \frac{(A(k_2, k_3)}{A(k_3, k_2)}\phi_1(k_1, k_3, k_2)\,,  \nn \\
&&\phi_5 = \frac{A(k_1, k_3)A(k_2, k_3)}{A(k_3, k_1)A(k_3, k_2)}\phi_1(k_3,
k_1, k_2)\,,  \nn \\
&&\phi_6 = \frac{A(k_2, k_3)A(k_1, k_3) A(k_1, k_2)}{A(k_3, k_2)A(k_3,
k_1)A(k_2, k_1)}\phi_1(k_3, k_2, k_1)\,.  \nn 
\eea
On the plane $x=0$, these $\phi_i$s' satisfy relations like 
\beq
\phi_2{|_{x=0}} = \frac{A(k_1, k_2)}{A(k_2, k_1)} \phi_1|_{x=0} \, ,
~~\phi_6|_{x=0} = \frac{A(k_2, k_3)}{A(k_3, k_2)} \phi_3|_{x=0} \, ,
~~\phi_5|_{x=0} = \frac{A(k_1, k_3)}{A(k_3, k_1)} \phi_4|_{x=0} \,.
\label {c5b}
\eeq
By using Eqs.(\ref{c5a}) and (\ref{c5b}), we obtain  
\bea
\phi_+|_{x=0} &=& \frac{2(k_1-k_2)}{(k_1-k_2) - i\xi(k_1 +
k_2)}\phi_1|_{x=0} + \frac{2(k_2-k_3)}{(k_2-k_3) - i\xi(k_2 +
k_3)}\phi_3|_{x=0} \nn \\
&&+ \frac{2(k_1-k_3)}{(k_1-k_3) - i\xi(k_1 +
k_3)}\phi_4|_{x=0}\,.
\label{c6}
\eea
Similarly, by differentiating $\phi_+$ in Eq.(\ref{c5a}) with respect 
to $x_3$  and  using Eq.(\ref{c5b}), we get 
\bea
\frac{\partial\phi_+}{\partial x_3}\Big|_{x=0}
&=&  \frac{2 \rho_1 (k_1-k_2)}{(k_1-k_2) - i\xi(k_1 +
k_2)}\phi_1|_{x=0} +  \frac{2 \rho_2 (k_2-k_3)}{(k_2-k_3) - i\xi(k_2 +
k_3)}\phi_3|_{x=0} \nn \\
&&+\frac{2 \rho_3 (k_1-k_3)}{(k_1-k_3) - i\xi(k_1 +
k_3)}\phi_4|_{x=0}\, ,
\label{c7}
\eea
where $\rho_1= -\frac{i}{3}(k_1+k_2-2k_3)$, $\rho_2= -\frac{i}{3}(k_2+k_3-2k_1)$ and 
$\rho_3= -\frac{i}{3}(k_1+k_3-2k_2)$. Comparing the r.h.s. of Eq.(\ref{c6}) 
with that of Eq.(\ref{c7}), we find that the condition (\ref{c4}) can only be satisfied 
if $\rho_1 = \rho_2 =\rho_3$, i.e., $k_1=k_2=k_3$. However, as is well known, 
all $k_i$'s take distinct values in the Bethe ansatz eigenfunction (\ref{c5}). 
Hence, these Bethe ansatz eigenfunctions do not satisfy the condition 
(\ref{c4}) for $N=3$. This in turn implies that the boundary condition 
(\ref {c3a}) can not be expressed
only through the partial derivative with respect
to the relative coordinate $x$. 
Therefore it is clear that, in contrast to what we have found 
 for $N=2$ case, the dual fermionic system associated with 
derivative $\delta$-function bose gas containing three particles 
 cannot be constructed by applying the method of self-adjoint extension to 
any short-range potential which depends on only one relative coordinate.
Proceeding in a similar way, one can 
arrive at the same conclusion for the case of  
derivative $\delta$-function Bose gas containing
more than three  particles. 


\noindent \section {Concluding Remarks}

In this article,
we attempt to construct the dual fermionic system associated with
the quantum integrable derivative $\delta$-function bose gas. At first, 
we consider the simplest case of two bosonic particles interacting 
through derivative $\delta$-function interaction. Similar to the
case of $\delta$-function potential, 
 the Hamiltonian of this two particle system can be reduced
to a single particle Hamiltonian depending only on the relative coordinate.
While this Hamiltonian preserves the continuity of the wavefunction,
it induces a discontinuity in the spatial derivative of the
wavefunction. On the other hand, the fermionic counterpart of this
bosonic wavefunction exhibits a discontinuity in the
wavefunction itself but not in its spatial derivative. By applying the method
of self-adjoint extension, we
construct a short-range potential which induces this type of discontinuity
in the wavefunction and yields the Hamiltonian of
the dual fermionic system. In contrast to the 
case of $\delta$-function bose gas, 
the coupling constant of this dual fermionic system becomes
explicitly dependent on the centre-of-mass momentum of the 
bosonic wavefunction.

Next we try to find out the dual fermionic model of the
derivative $\delta$-function bose gas for higher number of particles. 
Focussing on the case of derivative $\delta$-function bose gas with 
three particles, we find that the boundary condition on the corresponding 
eigenfunction can not be written
through a set of equations, each of which depends on the
partial derivative  with respect to only one relative coordinate.
The same conclusion can be drawn for the case of
derivative $\delta$-function bose gas containing
more than three  particles. Therefore, 
the dual fermionic system associated with
derivative $\delta$-function bose gas containing three 
or higher numbers of particles 
cannot be constructed in a similar way as we have done 
for the case of two particles.  To find out 
such dual fermionic system associated with 
higher numbers of particles, it seems to be necessary 
to study the self-adjoint extension of 
short-range potentials which depend on more than one variable.

\bigskip

\noindent {\bf Acknowledgments }

One of the authors (BBM) would like to acknowledge 
the Abdus Salam International Centre for Theoretical Physics for a Senior Associateship,
which partially supported this work. 




\newpage

\leftline {\large \bf References}
\medskip
\begin{enumerate}
\item M. Girardeau, J. Math. Phys. 1 (1960) 516; M. D. Girardeau, Phys. Rev.
B 139 (1965) 500. 


\item V. I. Yukalov,
M. D. Girardeau, Laser Phys. Lett. 2 (2005) 375. 

\item T Cheon, T. Shigehara, Phys. Rev. Lett. 82 (1999) 2536.  

\item A. G. Shnirman, B. A. Malomed, E. Ben-Jacob, Phys. Rev. A 50 (1994)
3453.

\item E. Gutkin, Ann. Phys. 176 (1987) 22.

\item A. Kundu, B. Basu-Mallick, J. Math. Phys. 34 (1993) 1052.

\item B. Basu-Mallick, T. Bhattacharyya, Nucl. Phys. B 634 (2002) 611.

\item B. Basu-Mallick, T. Bhattacharyya, Nucl. Phys. B 668 (2003) 415. 

\item  M. Wadati, H. Sanuki, K. Konno, Y.-H. Ichikawa, Rocky Mountain J.
Math. 8 (1978) 323; Y.-H. Ichikawa, S. Watanabe, J. de Physique 38 (1977)
C 6-15.

\item P. A. Clarkson, Nonlinearity 5 (1992) 453. 

\item 
M. Rosenbluh, R. M. Shelby, Phys. Rev. Lett. 66 (1991) 153.

\item U. Aglietti, L. Griguolo, R. Jackiw, S.-Y. Pi, D. Seminara, Phys. Rev.
Lett. 77 (1996) 4406; R. Jackiw, {\it A nonrelativistic chiral soliton in
one dimension,} hep-th/9611185.

\item S. J. Benetton Rabello, Phys. Rev. Lett. 76 (1996) 4007.

\item B. Basu-Mallick, T.Bhattacharyya, D. Sen, 
Nucl. Phys. B 675 (2003) 516; Phys. Lett. A 325 (2004) 375; 
 Mod. Phys. Lett. A 19 (2004) 2697.

\item F. Gesztesy, W. Kirsch, J. Reine Angew. Math. 362 (1985) 28.

\item P. \v{S}eba, Czech. J. Phys. B 36 (1986) 667.

\item P. \v{S}eba, Rep. Math. Phys. 24 (1986) 111.

\item S. Albeverio, F. Gesztesy, R. Hoegh-Krohn, H. Holden, {\it Solvable
Models in Quantum Mechanics} (Springer, Heidelberg, 1988).

\item T. Cheon, T. Shigehara, Phys. Lett. A 243 (1998) 111.

\item E. K. Sklyanin, in {\it Yang-Baxter Equation in Integrable Systems,}
World Scientific (Singapore) 1989. Advanced series in Math. Phys. Vol. 10,
p.121.

\item E. Lieb and W. Linger, Phys. Rev. 130 (1963) 1605.


\end{enumerate}

\end{document}